# Challenging theories of dark energy with levitated force sensor

Peiran Yin[1,2,7], Rui Li[2,3,4,7], Chengjiang Yin[5,6,7], Xiangyu Xu[5,6], Xiang Bian[1], Han Xie[1], Chang-Kui Duan[2,3,4], Pu Huang[1*], Jian-hua He[5,6*], Jiangfeng Du[2,3,4*]

[1] *National Laboratory of Solid State Microstructures and Department of Physics, Nanjing University, Nanjing, 210093, China*

[2] *CAS Key Laboratory of Microscale Magnetic Resonance and Department of Modern Physics, University of Science and Technology of China, Hefei 230026, China*

[3] *CAS Center for Excellence in Quantum Information and Quantum Physics, University of Science and Technology of China, Hefei 230026, China*

[4] *Hefei National Laboratory for Physical Sciences at the Microscale, University of Science and Technology of China, Hefei 230026, China*

[5] *School of Astronomy and Space Science, Nanjing University, Nanjing 210093, P. R. China*

[6] *Key Laboratory of Modern Astronomy and Astrophysics (Nanjing University), Ministry of Education, Nanjing 210093, China*

[7] *These authors contributed equally: Peiran Yin, Rui Li, Chengjiang Yin.*

*Corresponding author Email: hp@nju.edu.cn (P.H.); hejianhua@nju.edu.cn (J.H.); djf@ustc.edu.cn (J.D.)

**The nature of dark energy is one of the most outstanding problems in physical science, and various theories have been proposed[1]. It is therefore essential to directly verify or rule out these theories experimentally. However, despite substantial efforts in astrophysical observations[2-6] and laboratory experiments[7-20], previous tests have not yet acquired enough accuracy to provide decisive conclusions as to the validity of these theories. Here, using a diamagnetically levitated force sensor, we carry out a test on one of the most compelling explanations for dark energy to date, namely the Chameleon theory[21], an ultra-light scalar field with screening mechanisms, which couples to normal-**

**matter fields and leaves a detectable "fifth force". Our results extend previous results by nearly two orders of magnitude to the entire physical plausible parameter space of cosmologically viable chameleon models. We find no evidence for such a "fifth force". Our results decisively rule out the basic chameleon model as a candidate for dark energy. Our work, thus, demonstrates the robustness of laboratory experiments in unveiling the nature of dark energy in the future. The methodology developed here can be further applied to study a broad range of fundamental physics[22-25].**

On cosmic scales, a mysterious component with positive energy but negative, repulsive pressure makes up the bulk of the Universe, which drives the expansion of the Universe to speed up[26,27]. This exotic component is called dark energy. One explanation for dark energy is Einstein's cosmological constant, which is potentially related to the vacuum energy of quantum fields. However, current quantum field theories can neither naturally predict the measured small value of the cosmological constant nor explain its stability without fine-tuning.

Alternatively, dark energy can be explained as a dynamic scalar field. One compelling example is the so-called chameleon field[21] $\phi$, whose equation-of-motion (using natural units) is

$$\nabla^2 \phi = \frac{\partial V_{\text{eff}}}{\partial \phi} \qquad (1)$$

with $\nabla^2$ being the Laplace operator, and $V_{\text{eff}}$ the effective potential

$$V_{\text{eff}} = \Lambda^4 \left(1 + \frac{\Lambda^n}{\phi^n}\right) + \rho \left(1 + \frac{\phi}{M_\beta}\right) \qquad (2)$$

The first term in $V_{\text{eff}}$ indicates self-interaction and the second term describes the interaction with ordinary matter of density $\rho$. For cosmologically motivated chameleon, the coupling between $\phi$ and normal-matter field is characterized by an energy scale $M_\beta = M_{\text{Pl}}/\beta$, which is expected below the reduced Planck mass $M_{\text{Pl}} = (\hbar c/8\pi G)^{1/2} \approx 2.4 \times 10^{18}$ GeV/c$^2$, with $\beta$ a dimensionless factor. The energy scale $\Lambda$ is usually taken to be close to the cosmological-constant, *i.e.*, $\Lambda \approx \Lambda_0 = 2.4$ meV, which accounts for the cosmic acceleration. The real index $n$ is often taken to be 1 (referred to as the basic chameleon model hereafter). The non-linear term with inverse power of $\phi$ leads to a screening mechanism, which depends on the ambient matter density. In sparse environments, such as the cosmos, the chameleon field is light and mediates a long-range "fifth force" but in a high-density environment, such as the laboratory, $\phi$ becomes massive and the "fifth force" is suppressed.

Despite as a compelling theory for dark energy, a precise test of the "fifth force" of the chameleon field is indeed challenging. Although arguably less effective is the screening at large scales, cosmological and astrophysical observations[2-6] are prone to systematic errors. This hinders a robust test based on those observations. Laboratory experiments, however, are facing the problem of double suppression. The "fifth force" is screened not only at the source masses but also at the force sensor, which makes it difficult to be detected. Although the precision of torque in torsion pendulum experiments[7-10] is very high, when it comes to testing the "fifth force", it becomes less effective when the screening is strong. To alleviate such double suppression, atom interferometry has been proposed[11], in which atoms are used as the force sensors. Due to their smallness, the self-screening of atoms can be neglected. Significant improvements, thus, are obtained[12-14]. However, the reported atom interferometry experiments are beset by the small size of source masses, which leads to a limited useful free-fall time for particles to sample the chameleon field. As a result, despite the progresses that have been made in the past decade, a key region in the parameter space from $M_\beta \sim 10^{-3}\ M_{\mathrm{Pl}}$ to $M_\beta \sim 10^{-1}\ M_{\mathrm{Pl}}$, which allows for cosmological viable chameleons, still has not been covered. This region spans more than two orders of magnitude, to fill which requires significant improvement of techniques in existing experiments. It, therefore, remains a challenge for state-of-the-art laboratory experiments.

Here, we report a test on such a "fifth force" by taking a different route. We adopt a diamagnetically levitated force sensor (see Fig.1a, b), a system that has been emerging to be ultra-sensitive for force detection at sub-milligram scales[28,29]. We use a thin-film structure particularly designed for both the force sensor and source masses to overcome the problem of double suppression. The geometries of the force sensor and source masses are carefully optimized to maximize the produced "fifth force" by employing numerical simulations (see Methods for details). Moreover, we manage to generate a long-time coherent "fifth force", which can significantly improve the force detectability.

Figure 1a shows the schematic of our experiment. Eight thin films of polyimide spaced equally on a rotating plate are used as the source masses to generate a periodic chameleon field. The field then penetrates a vacuum chamber via a thin window and exerts the "fifth force" on a force sensor suspended inside the chamber. The force sensor consists of a thin film of polyimide at the top supported by a glass rod and a piece of pyrolytic graphite at the bottom.

The pyrolytic graphite works as a supporter and is levitated in the magneto-gravitational trap via diamagnetic force (see Methods, Extended Data Fig. 4). The thin film is used as a test mass, which is the part of the force sensor that can effectively feel the periodic "fifth force". This is because the "fifth force" below the thin film is screened by a magnetic shielding box that encloses the pyrolytic (see Methods). To enhance the detectability of the "fifth force", we choose the thin films with a large surface area and optimize the thickness as 12.5μm, which is comparable to the Compton wavelength of the chameleon in the parameter space of interest. We take the length of the glass rod long enough that the thin film is placed close to the source mass. In practical measurements, the drive frequency $\omega_{dri}/2\pi$ is set at the resonance frequency $\omega_0/2\pi$ of the force sensor along the $z$-direction. The motion of the force sensor is monitored optically, as shown in the inset of Fig. 1a. Finally, the magneto-gravitational trap is placed on a vibrational-free stage in vacuum.

In addition to the chameleon "fifth force", in practice, however, the detected forces may contain background contributions, such as magnetic, electrostatic forces as well as the Newtonian gravity of the source masses. Therefore, to make a clean test of the chameleon field, it is important to effectively mitigate these effects. For the magnetic forces, we use a magnet shielding box that encloses the magneto-gravitational trap as well as the pyrolytic graphite. Only a small hole is left at the top, which allows the test mass to get out of the shielding box (see Fig. 1b). We find that such a scheme works very well in suppressing the magnetic forces in our experiments (see Methods, Extended Data Fig. 7). For the electrostatic forces, the vacuum chamber itself, indeed, is an ideal Faraday shielding cage, thanks to the design that the source masses on the rotating plate are separated outside the chamber (see Methods, Extended Data Fig. 8). However, the chamber walls also hinder the "fifth force" from passing through and acting on the test mass. Therefore, to mitigate such effects, we make a metalized low strain silicon nitride window as thin as 0.5 μm right below the source masses at the top of the chamber (see Fig. 1b). Finally, the thickness of thin films of the source masses is chosen as 75 μm, and the Newtonian gravity produced by the source masses is over two orders of magnitude weaker than the expected chameleon "fifth force" in the parameter space of interest (see

Methods, Extended Data Fig. 9). Figure 1c plots the potential of the chameleon field $\phi$ along the central $z$-axis of both cases, with and without a film of source mass above the test mass. This shows that different $\phi$ is produced at different rotation phases, which will lead to different "fifth force" at the test mass.

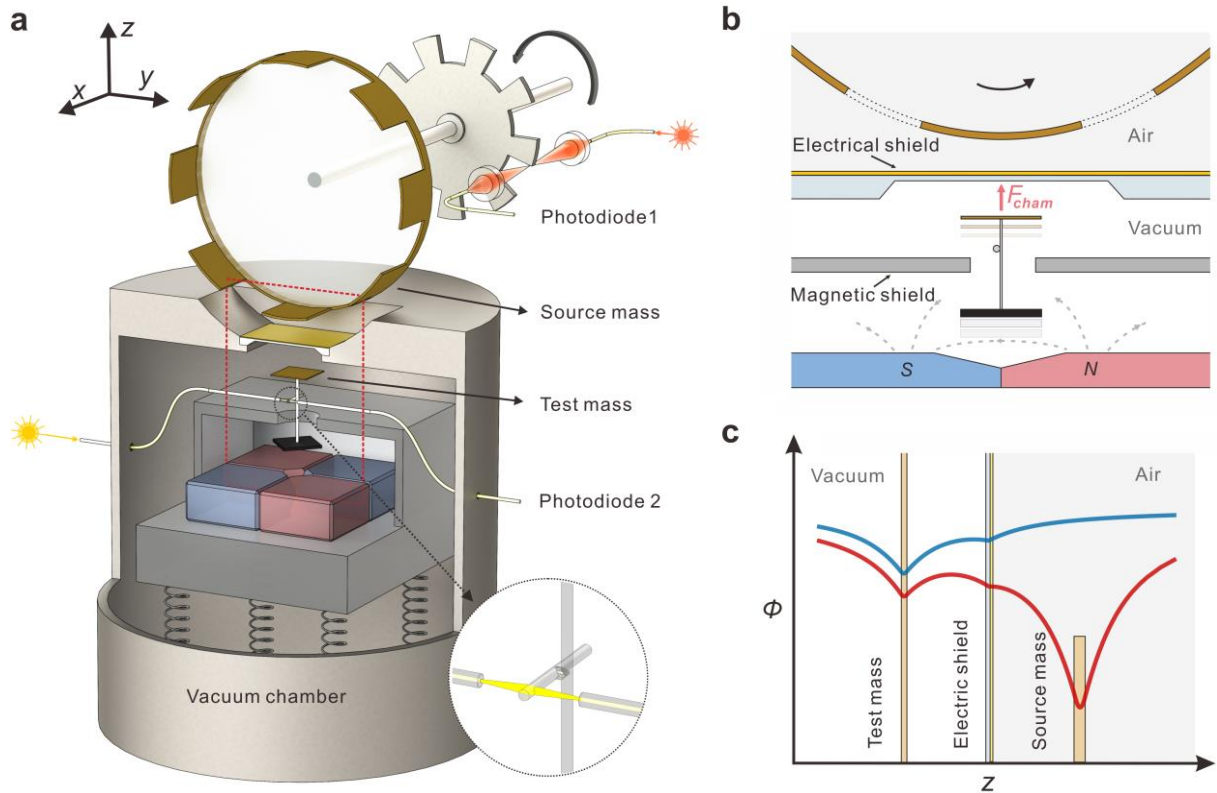

**Fig. 1 | Schematic of experiment. a**, The "fifth force" of the chameleon field is generated by 8 thin films (source masses) of polyimide with a thickness of 75 μm, which are spaced equally on a rotating plate. The force sensor consists of a piece of pyrolytic graphite diamagnetically levitated in a magneto-gravitational trap and a 12.5 μm-thick film with the same material as the source masses at the top supported by a glass rod. The magneto-gravitational trap is placed in a vacuum chamber with seismic noise isolation. The distance between the test mass and the source masses is about 400 μm. The rotation of source masses and motion of the force sensor are monitored by optical systems. **b**, The rotating source masses generate a periodic "fifth force" acting on the test mass. A thin electrical shielding window with a thickness of 0.5 μm and a magnetic shield are used to screen the background electrostatic and magnetic forces. **c**, The field $\phi$ along the central $z$-axis at two different rotation phases. The red and blue curves

indicate the cases with and without a film of source mass above the test mass, respectively.

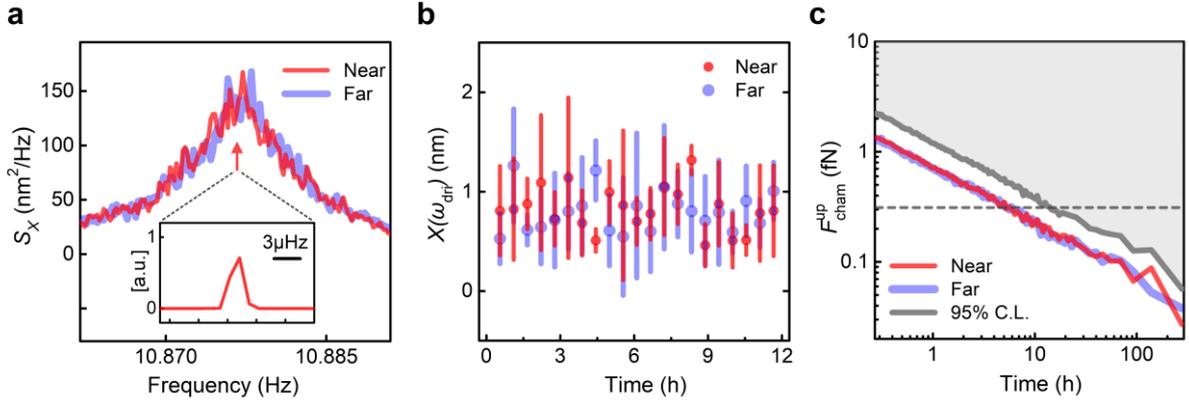

**Fig. 2 | Experiment data. a**, Power spectral density of the displacement of the force sensor, the red and blue curves represent the experiments with source masses at near and far positions, respectively. The drive frequency of source masses is $\omega_{\mathrm{dri}}/2\pi = \omega_0/2\pi$ as indicated by the red arrow and the corresponding power spectral density is shown in the inset. **b**, A 12-hour section of data. The displacement response $X(\omega_{\mathrm{dri}})$ for the two runs with source masses at near and far positions, respectively. Each point represents a 400 s measurement. Error bars are derived from five repetitions of measurements. **c**, The upper bound of the chameleon "fifth force" as a function of measurement time at the 95% confidence level. The gray shaded region shows the magnitude of the excluded "fifth force" as a function of measurement time. The gray dashed line indicates the "fifth force" predicted by the basic chameleon model at the dark energy scale $\Lambda = 2.4$ meV, $M_\beta = 10^{-2} M_{\mathrm{Pl}}$.

In practical measurements, besides the background forces that are coherent with the "fifth force" (see Methods for details), the incoherent thermal Brownian noises from the environment are also applied to the force sensor at a finite temperature. To subtract such incoherent noises, we perform two independent runs. In the first run, we place the source masses at a position with the closest distance between the source masses and the test mass about 400 μm estimated by optical observation (denoted as near position, see Methods). The displacement response $X(t)$ of the force sensor is recorded optically. The "fifth force", if any, is at its maximum in this case. The second run works as a control, in which all the parameters are kept the same as the first run but with the source masses being moved away from the test mass to about 3 cm (denoted as far position). The "fifth force", in this case, is many orders of magnitudes weaker than that in the first run. As such, the displacement response of the force sensor $X(t)$ only arises from the thermal Brownian noises.

The corresponding displacement response $X(\omega_{\text{dri}})$ at the drive frequency is calculated by

$$X(\omega_{\text{dri}}) = 1/t \left| \int X(t) e^{i\omega_{\text{dri}} t} \, dt \right| \tag{3}$$

where $t$ is the measurement time. The power spectral density of $X(t)$, which is defined by $S_X = \langle t\, X^2(\omega) \rangle$, is shown in Fig. 2a. The source masses are driven by a servo motor with high stability in frequency and the "fifth force" is treated as ideally periodic at frequency $\omega_{\text{dri}}/2\pi = \omega_0/2\pi$ in our analysis (see Methods for details). Figure 2b shows the typical data of $X(\omega_{\text{dri}})$ within 12 hours for the two experiment runs. Each point represents a 400 s measurement averaged 5 times. The displacement response of the drive force $X^2_{\text{dri}}(\omega_{\text{dri}})$ is then calculated by $\langle X^2_{\text{dri}}(\omega_{\text{dri}}) \rangle = \langle X^2_{\text{near}}(\omega_{\text{dri}}) \rangle - \langle X^2_{\text{far}}(\omega_{\text{dri}}) \rangle$, where $\langle X^2_{\text{near}}(\omega_{\text{dri}}) \rangle$ and $\langle X^2_{\text{far}}(\omega_{\text{dri}}) \rangle$ indicate the measurements at the near and far positions, respectively. From the data, we estimated the upper bound of the "fifth force" through the mechanical response function of a harmonic oscillator (see Methods). Figure 2c plots the upper bound of the "fifth force" as a function of measurement time at the 95% confidence level. Given the two 11.6-day measurements, the upper bound is estimated as $5.7 \times 10^{-17}$ N. The gray shaded region shows the magnitude of the excluded "fifth force" as a function of measurement time. The gray dashed line indicates the "fifth force" predicted by the basic chameleon model at the dark energy scale $\Lambda = 2.4$ meV, $M_\beta = 10^{-2} M_{\text{Pl}}$.

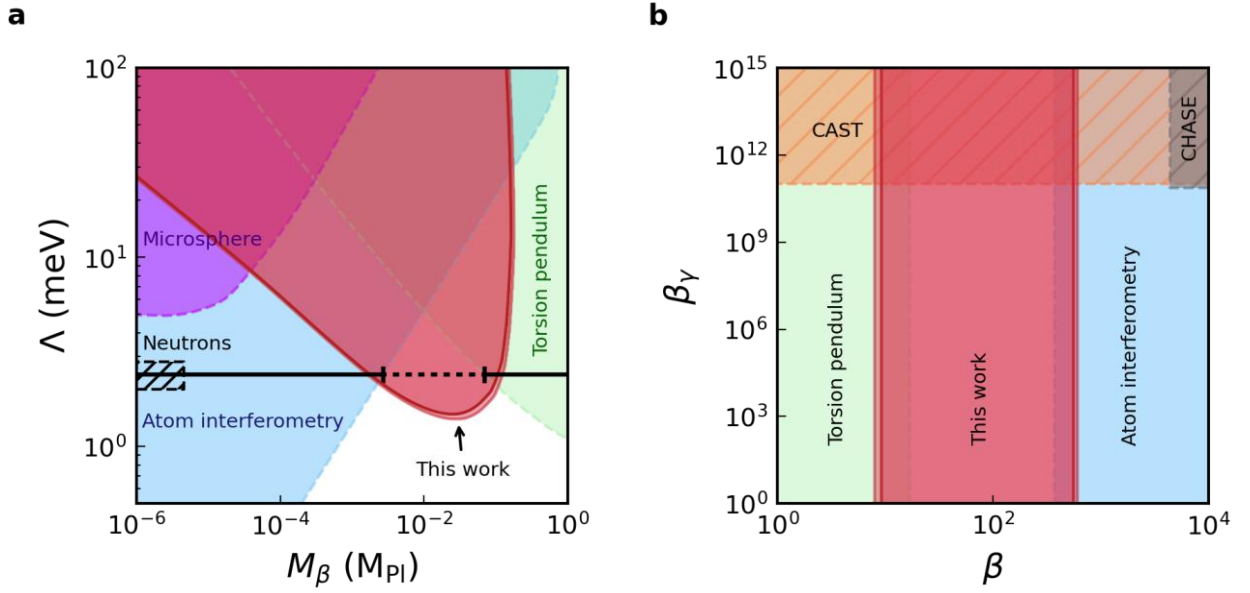

**Fig. 3 | Exclusion of the chameleon field. a**, $\Lambda - M_\beta$ plane for the basic chameleon model ($n = 1$). The red shaded region is ruled out by our experiments at a 95% confidence level. The darker and lighter layers indicate the differences considering the systematic uncertainty in our experiments (see Methods, Extended Data Table 1). Regions that were excluded by previous neutron experiments[15-18] and microsphere force sensing[19] are also shown. The solid black line segments represent the excluded regions at the dark energy scale $\Lambda = 2.4$ meV by previous torsion pendulum experiments[7-9] and atom interferometry[12-14]. Our results completely fill the gap between them (dashed black line segment). **b**, Comparison with CHASE[20] and CAST[5] experiments that assume photon coupling $\beta_\gamma$. In combination with atom interferometer and torsion balance experiments, our results rule out the coupling $\beta = M_{Pl}/M_\beta$ between normal-matter and the basic chameleon model.

Specializing to the chameleon fields, Figure 3a shows the excluded parameter space of the basic chameleon model in the $\Lambda - M_\beta$ plane. The red shaded regions show the parameter space excluded by our experiments at the 95% confidence level. For cosmologically viable chameleons with $\Lambda = 2.4$ meV, our results exclude the energy scale $M_\beta$ in the range of $1.6 \times 10^{-3} M_{Pl} < M_\beta < 1.2 \times 10^{-1} M_{Pl}$, completely filling the gap between torsion pendulum experiments[7-9] and atom interferometry constraints[12-14]. For $\Lambda > 2.4$ meV, the parameter space of all such models is completely ruled out. Figure 3b compares our results with experiments assuming an additional coupling between the chameleon field and the photons beyond the fifth force[5,20]. Our results fill the gap and rule out the coupling between normal-matter and the basic chameleon model ($n = 1$ and $\Lambda = 2.4$ meV).

In conclusion, we have made a precise test on the basic chameleon model for the energy scale $M_\beta$ in the range of $1.6 \times 10^{-3} M_{\text{Pl}} < M_\beta < 1.2 \times 10^{-1} M_{\text{Pl}}$ ($\Lambda = 2.4$ meV). We find no evidence for such a "fifth force". Our results rule out the cosmologically viable basic chameleon model, which leads to a decisive conclusion that such models cannot be a solution to the conundrum of cosmic acceleration. Our platform can be generalized to test other theories, such as symmetron[30] and $f(R)$ theories[31]. Moreover, our methodology can be generalized to other levitated systems[32-36]. Our work demonstrates that besides the conventional large cosmological projects, such as the ground-based Dark Energy Spectroscopic Instrument (DESI) project[37] and the space-borne Euclid mission[38], laboratory experiments can provide an alternative and promising way for studying dark energy, which may unveil the nature of dark energy in the future. Finally, the performance of our experiment can also be significantly improved in a cryogenic environment[39]. As such, our system can be developed for studying a wide range of fundamental physics problems such as short-range gravity[22,23], the mechanism of wave-function collapse[24] and the quantum gravity[25].

## Methods

## Theory

We solve the non-linear equation of the chameleon field using numerical methods. Our approach is based on the finite element method (FEM) as well as Newton's method. Compared to other conventional methods, such as the finite difference method (FDM), the FEM can be applied to irregular meshes, which is well-suited for systems with complex geometries and boundaries. Moreover, the FEM can be massively parallelized in computation, which allows for a high-resolution calculation of the field profile on a three-dimensional grid, covering the entire details of our apparatus.

**Chameleon equation normalization**

From Eq. (1) in the main text, the equation of motion of the chameleon field is given by

$$\nabla^2 \phi = \frac{\partial V_{\text{eff}}}{\partial \phi} = -\frac{n\Lambda^{n+4}}{\phi^{n+1}} + \frac{\rho}{M_\beta} \tag{1}$$

Due to the screening effect, the field gradient vanishes rapidly, *i.e.*, $\nabla^2 \tilde{\phi} \approx 0$, in a large enough homogenous background with density $\rho_{\text{bg}}$, and thus $\phi$ reaches equilibrium. We denote such an equilibrium value as

$$\phi_{\text{bg}} \equiv \phi_{\text{eq}}(\rho_{\text{bg}}) = \left(\frac{nM_\beta \Lambda^{n+4}}{\rho_{\text{bg}}}\right)^{\frac{1}{n+1}} \tag{2}$$

In the numerical process, it is more convenient to work on a normalized equation with $\tilde{\phi} = \frac{\phi}{\phi_{\text{bg}}}$ and $\tilde{\rho} = \frac{\rho}{\rho_{\text{bg}}}$, then we have

$$\nabla^2 \tilde{\phi} = -\frac{n\Lambda^{n+4}}{\tilde{\phi}^{n+1} \phi_{\text{bg}}^{n+2}} + \frac{\tilde{\rho}\rho_{\text{bg}}}{M_\beta \phi_{\text{bg}}} \tag{3}$$

Let $r_{\text{bg}}$ be the Compton wavelength[42] of the background field

$$r_{\text{bg}}^2 = \frac{\phi_{\text{bg}}^{n+2}}{n(n+1)\Lambda^{n+4}} = \frac{M_\beta \phi_{\text{bg}}}{(n+1)\rho_{\text{bg}}} \tag{4}$$

Equation (2) can be rewritten as

$$\nabla^2 \tilde{\phi} = -\frac{1}{(n+1)r_{bg}^2}\left(\tilde{\phi}^{-(n+1)} - \tilde{\rho}\right) \tag{5}$$

Compared with Eq. (1), $\tilde{\phi}$ in Eq. (5) only explicitly depends on $r_{bg}$ rather than $\Lambda$ and $M_\beta$, once the power index $n$ and the density profile $\tilde{\rho}$ are given. As such, to explore the parameter space of the chameleon model, we only need to solve $\tilde{\phi}$ for different $r_{bg}$ first and then use Eq. (2) to rescale back to $\phi$ with arbitrary $\Lambda$ and $M_\beta$ along the iso-Compton wavelength lines, where $r_{bg}$ is fixed and $\Lambda \propto \beta^{\frac{n+2}{n+4}}$.

**Newton's method and spatial discretization**

Since Eq. (5) is a non-linear equation, we adopt Newton's method to linearize and solve it iteratively. For a $k$-th approximate solution $\tilde{\phi}_k$, the update term $\delta\tilde{\phi}_k$ is given by a linear equation

$$L'(\tilde{\phi}_k)\delta\tilde{\phi}_k = -L(\tilde{\phi}_k) \tag{6}$$

where

$$L(\tilde{\phi}_k) = \nabla^2\tilde{\phi}_k + \frac{1}{(n+1)r_{bg}^2}\left(\tilde{\phi}_k^{-(n+1)} - \tilde{\rho}\right) \tag{7}$$

and the prime denotes the derivatives with respect to $\tilde{\phi}_k$. Equation (6) then becomes

$$\nabla^2\delta\tilde{\phi}_k - \frac{\tilde{\phi}_k^{-(n+2)}}{r_{bg}^2}\delta\tilde{\phi}_k = -\nabla^2\tilde{\phi}_k - \frac{1}{(n+1)r_{bg}^2}\left(\tilde{\phi}_k^{-(n+1)} - \tilde{\rho}\right) \tag{8}$$

Once the above equation is solved, the $(k+1)$-th approximate solution $\tilde{\phi}_{k+1}$ can be updated by

$$\tilde{\phi}_{k+1} = \tilde{\phi}_k + \alpha_k \delta\tilde{\phi}_k \tag{9}$$

where $\alpha_k$ is step length. In Newton's method, once an initial guess $\tilde{\phi}_0$ is given, the approximate solution $\tilde{\phi}_{k+1}$ can be iteratively obtained until it converges to the actual solution of Eq. (5). In practice, the convergence of $\tilde{\phi}_{k+1}$ is quantified via the residual $L(\tilde{\phi}_k)$ being less than a stringent criterion.

In this work, the FEM is applied to solve Eq. (8). We first put it into its variational form, in which the equation is multiplied by a set of test functions $\{\varphi^i\}_1^N$ and then integrated over the simulation domain. Equation (8) then becomes

$$-\langle \nabla \varphi^i, \nabla \delta \tilde{\phi}_k \rangle - \langle \varphi^i, \frac{\tilde{\phi}_k^{-(n+2)}}{r_{bg}^2} \delta \tilde{\phi}_k \rangle = \langle \nabla \varphi^i, \nabla \tilde{\phi}_k \rangle - \langle \varphi^i, \frac{1}{(n+1)r_{bg}^2} \left( \tilde{\phi}_k^{-(n+1)} - \tilde{\rho} \right) \rangle \quad (10)$$

where the brackets denote $\langle f, g \rangle = \int f \cdot g \, dx$ for convenience. If Eq. (10) holds for any test function $\varphi^i$, $\delta \tilde{\phi}_k$ is then called the variational solution of Eq. (8).

In practice, however, it turns out that directly solving for $\delta \tilde{\phi}_k$ may cause instabilities during iterations. To avoid this problem, we employ the transformation $e^{u_k} = \tilde{\phi}_k$ instead. This new variable $\delta u_k$ relates to $\delta \tilde{\phi}_k$ via $\delta \tilde{\phi}_k = e^{u_k} \delta u_k$ and $\nabla \delta \tilde{\phi}_k = e^{u_k} \nabla u_k \delta u_k + e^{u_k} \nabla \delta u_k$. Further, $\delta u_k$ can be expanded by the test functions

$$\delta u_k = \sum_{j=1}^N \delta U_k^j \varphi^j \quad (11)$$

where $\delta U_k^j$ are coefficients to be solved. Inserting the above expressions back into the variational formula, we finally obtain a linear system

$$\sum_{j=1}^N \left[ -\langle \nabla \varphi^i, e^{u_k} (\nabla \varphi^j + \varphi^j \nabla u_k) \rangle - \langle \varphi^i, \frac{e^{-(n+2)u_k}}{r_{bg}^2} \varphi^j \rangle \right] \delta U_k^j$$

$$= \langle \nabla \varphi^i, e^{u_k} \nabla u_k \rangle - \langle \varphi^i, \frac{1}{(n+1)r_{bg}^2} \left( e^{-(n+1)u_k} - \tilde{\rho} \right) \rangle, \quad i = 1,2,3,\ldots,N \quad (12)$$

from which $\delta U_k^j$ can be solved.

**Numerical results**

The number of independent equations in the linear system Eq. (12), also known as degree of freedom (DoF), is usually very large, which can be easily up to $10^9$. Hence, direct linear solvers such as the LU decomposition are inefficient and one needs to turn to the iterative solvers. However, since the matrix for $\delta U_k^j$ is not symmetric in our case, the classic conjugate gradient (CG) method cannot be applied. Instead, we use the GMRES method[40], which does not require any specific properties of the matrices.

In addition, we use a method of line-search to optimize the step length $\alpha_k$ for each step. We choose $\alpha_k$ in such a way that it gives the minimum residual of $L(\tilde{\phi}_k)$ along the direction of $L'(\tilde{\phi}_k)$. We find that this varying $\alpha_k$ can significantly improve the stability and convergence rate of Newton's method.

In practice, our numerical implementation is based on the open source FEM library deal.II[41], which is written in C++. The code supports massively parallelization and local adaptive refinement.

Extended Data Fig. 1 shows the numerical tests of our code. In these tests, we assume a uniform sphere of 1mm in radius is surrounded by air of a large enough volume. The density of the sphere is chosen to be $10 \text{ kg/m}^3$ and $100 \text{ kg/m}^3$ to illustrate weak and strong screening case, respectively. For the low-density test $\rho = 10 \text{ kg/m}^3$, we take the Compton wavelength of the sphere as $r_c = 0.27$ mm ($\Lambda = 2.4$ meV, $\beta = 100$), which is comparable to its radius. The blue curve shows the field profile from our calculation. Since in this case, the screening effect is relatively weak, the potential deep inside the sphere does not reach the equilibrium value $\phi_{\text{eq}}$. However, for the high-density test $\rho = 100 \text{ kg/m}^3$, the same parameter set gives $r_c = 0.05$ mm, which is much smaller than the radius. In this strong screening case (red curve), the field reaches $\phi_{\text{eq}}$ quickly as expected, leaving the signals of the field only within a thin shell near the outer regions. Outside the sphere, both fields increase gradually and reaches $\phi_{\text{eq}}$ in the air. The black curve shows the theoretical prediction of the field profile under the strong screening approximation. Our numerical results closely match the theoretical approximate solution[42] (solid black curve).

Extended Data Fig. 2a shows a two-dimensional slice of grid points used in our calculation.

The color-coded dots represent the matter density of the instrument. Note that the grid points are not uniformly refined, which allows us to have sufficiently high resolution in regions of interest while keeping the overall computation cost at a moderate level. The inset shows the grids around the electrical shielding membrane, where the finest resolution is about 0.5 μm. The thin film of the force sensor, which works as the test mass, is well resolved in our simulations as well.

In practice, the boundary conditions are set as the homogeneous Dirichlet boundary conditions. Given the fact that the actual instrumental size is much larger than the Compton wavelength in the parameter space of interest ($r_{\text{air}} = 0.2 \sim 10$ mm), the background density is chosen as the density of the air $\rho_{\text{bg}} = 1.29 \text{kg/m}^3$ and $\phi_{\text{bg}}$ is taken as the equilibrium value of the air. The total DoFs in our simulation is about $1.5 \times 10^9$. The simulations use 960 CPU cores and are performed at the High-Performance Computing Center (HPCC) of Nanjing University.

Extended Data Fig. 2b, c show the numerical calculation of the chameleon field $\phi$ around the test mass of the force sensor in our experiment. The left panel shows the distribution of $\phi$ at two different rotation phases of the plate $\theta$ with (left half, $\theta = 0$) and without the source masses (right half, $\theta = \pi/8$). The right panel shows the field along the central z-axis with phases varying from $\theta = 0$ (blue curves) to $\theta = \pi/8$ (red curves). The distance between the test mass and the source mass is optimized to maximize the variation of $\phi$ at the force sensor between different phases.

After obtaining the chameleon field $\phi$, the "fifth force" felt by the force sensor can be calculated[42] by integrating the test mass region with density $\rho_{\text{film}}$ and volume $V_{\text{test}}$. Noting that the instrument is symmetric on the $xy$-plane, only the "fifth force" along the $z$-axis is needed:

$$F_{\text{cham}}|_z = -\frac{1}{M_\beta} \iiint_{V_{\text{test}}} \rho_{\text{film}} \nabla \phi \cdot \hat{z} dV = -\frac{\phi_{\text{bg}} \rho_{\text{film}}}{M_\beta} \iiint_{V_{\text{test}}} \nabla \tilde{\phi} \cdot \hat{z} dV = \frac{\phi_{\text{bg}} \rho_{\text{film}}}{M_\beta} \iint (\tilde{\phi}|_{z_1} - \tilde{\phi}|_{z_2}) dA \quad (13)$$

where $\tilde{\phi}|_{z_1}$ and $\tilde{\phi}|_{z_2}$ are the normalized fields on the upper and lower surfaces of the test mass. As the fields become indistinguishable when approaching the magnetic shielding box, we have tested that the contribution of other parts of the force sensor below the test mass to $F_{\text{cham}}$ is negligible.

We decompose $F_{\text{cham}}$ felt by the force sensor in terms of a normalized phase function $f(\theta)$ with the peak-to-peak value $F_{\text{cham}}^{\text{pp}}$ and an offset $F_{\text{cham}}^{\text{off}}$

$$F_{\text{cham}} = F_{\text{cham}}^{\text{pp}} f(\theta) + F_{\text{cham}}^{\text{off}} \qquad (14)$$

Compared with Eq. (13), $f(\theta)$ should only depend on $r_{\text{bg}}$. Extended Data Fig. 3 plots $f(\theta)$ in one period of $\pi/4$ for different parameter sets. These curves are very close to one another within the parameter space of our interest. Hence, it is more convenient to use an averaged phase curve template $f(\theta)$ to get a conversion factor $\eta$, with which $F_{\text{cham}}^{\text{pp}}$ can be converted directly to the power spectral density on the resonance peak (as discussed below), rather than solving the whole phase curve from the oscillation equation for every parameter set. This could significantly reduce the computational costs and the uncertainty induced is less than $\lesssim 0.5\%$.

**EXPERIMENT**

**Experimental system and parameter uncertainty**

The force sensor consists of a piece of thin film made of polyimide at the top and a piece of pyrolytic graphite at the bottom (see Extended Data Fig. 4). A 5 mm-long and 50 μm-diameter glass rod is attached vertically by UV glue on the pyrolytic graphite, which is denoted as the support rod. The thin film used as test mass is then attached to the top of this support rod. A short glass rod is further attached horizontally to the support rod, worked as a detection rod for measuring the motion of the force sensor.

In practice, the measurement of the motion of force sensor is made via an optical scheme. An incident laser light is illuminated on the detection rod through a fiber. A second fiber is placed behind the detection rod to collect the unobstructed light (see Fig. 1 in the main text). As such, the changes in the displacement of the force sensor are measured through the change in the intensity of the laser[28]. In our experiment, the amounts of UV glue are used as small as possible to minimize the total mass of the force sensor. The total mass of the sensor is measured using an analytical balance. We repeated the measurements several times to build up statistics. Moreover, we adopt the magneto-gravitational trap with a commonly used geometric structure. The sensor is then carefully put on it and is levitated automatically. In practice, we made many such force sensors with slightly different sizes in geometry. We choose the one with the lowest mechanical dissipation to minimize the effect of thermal Brownian noises.

The source masses consist of 8 thin films of polyimide spaced equally on a non-magnetic plate with a radius of 25 mm. The thin films are separated by 22.5-degrees from one another. The number of thin films is, indeed, carefully chosen to avoid the potential excitations of high harmonics from the servo motor. We choose the thin film as 3 mm wide. The thickness of the film is chosen at an optimal value of 75 μm. The rotating plate is driven by a servo motor through a rotary shaft. A gear-shaped plate with 8 equally spaced gaps is fixed on the rotary shaft, which is used to monitor the rotation of the source masses through an independent optical system.

The distances from the electrical shielding window to the film of force sensor (denoted as $d_1$) and the films of source masses (denoted as $d_2$) are tuned with a homemade positioner, and is directly measured using optical microscope with imprecision of about 20 μm.

In an ideal case, the film of force sensor should be parallel to the electrical shielding window and the surface of the films of the source masses. In practice, however, this cannot be fully realized. There will be some misalignments denoted as angles of rotation $\alpha_x$, $\alpha_y$, and $\alpha_z$ around the $x$, $y$ and $z$-axis between the sensor film and electrical shielding window. We estimate such uncertainties through optical observation. Other uncertainties, such as the relative position between the force sensor and the source masses in the horizontal plane, the density, and thickness of the film of the force sensor, source masses, electrical shield, and so on, are negligible.

The seismic isolation system consists of a two-stage spring-mass based suspension. The vibration noise from environment is well suppressed bellow the thermal noise of the mechanical oscillator. The mechanical parameters of the experimental system are summarized in Extended Data Table 1.

**Linewidth of the diamagnetically levitated oscillator and the rotating source masses**

For a harmonic oscillator, the linewidth is defined as the Lorentz-fitting full width at half maximum (FWHM) of power spectral density (PSD) of oscillator displacement, which can be expressed as $\Delta\omega/2\pi$. For an ideal oscillator, we have $\Delta\omega = \gamma$, where $\gamma$ is the mechanical dissipation. In a practical system, the measured linewidth is affected by factors such as temperature fluctuations, which may lead to a drift in resonance frequency and broaden the linewidth. In our experiment, the linewidth $\Delta\omega/2\pi$ of the force sensor is obtained by fitting the measured PSD of the oscillator displacement $S_X(\omega)$ to a Lorentz function (see. Fig. 2a in the main text). The value we obtained is $\Delta\omega/2\pi = 9.9$ mHz. The drift in resonant frequency due to the temperature fluctuations is denoted as $\delta\omega_0/2\pi$. The temperature of the sample stage that supports the magnets is controlled using the proportional-integral-derivative (PID) method with a standard deviation on the temperature of about 8mK. The frequency drift is carefully monitored during the entire experiment. Extended Data Fig. 5a shows the distribution of resonant frequency drift $\delta\omega_0/2\pi$ measured every 4000 s. We fit the measurement to a Gaussian distribution $P(\delta\omega_0) \sim e^{-(\delta\omega_0)^2/2\sigma^2}$. The statistic fluctuation obtained is $\sigma/2\pi = 0.91$ mHz, which is much smaller than the measured linewidth. In addition to temperature, the nonlinearity of the magnetogravitational trap affects the linewidth only when the amplitude of motion is large[43]. For the thermal Brownian motion, such effects are negligible. Therefore, the measured linewidth is used to determine the mechanical dissipation $\gamma$. It is worth noting that the measured linewidth in this experiment is much larger than the typical value in our previous experiment, where a diamagnetically levitated oscillator with a similar mass was used[28]. The reason here is that the material used in this experiment is pyrolytic graphite, which is conductive. Hence, strong magnetic damping due to the eddy current broadens the linewidth[44]

Another important issue is the linewidth of source masses, which are driven by a servo motor. In an ideal condition, the frequency of the "fifth force" at time $t$, indicated as $\omega_{\text{dri}}(t)/2\pi$, is constant and the linewidth is infinitely small. While the instabilities of the servo motor may lead to random fluctuations in frequency $\delta\omega_{\text{dri}}/2\pi$. In our experiment, we used an independent optical monitor to record the rotation of source masses. The output laser intensity $I(\theta(t))$ of the rotation monitor is normalized and has the form of a square wave with a period of $\pi/4$:

$$i(\theta(t)) = \begin{cases} 1, & 0 \le \theta(t) < \pi/8 \\ 0, & \pi/8 \le \theta(t) < \pi/4 \end{cases} \tag{15}$$

Here the normalization factor $\epsilon = 0.0155$ is used for convenience, $I(\theta(t)) = \epsilon i(\theta(t))$. Since there are 8 pieces of source films on the rotating plate, the drive frequency $\omega_{\text{dri}}/2\pi$ is eight times of the rotating frequency of the plate, that the rotation phase of the plate is expressed as $\theta(t) = \int \frac{1}{8}\omega_{\text{dri}}(t)dt$. As shown in Fig. 2a in the main text, the FWHM of PSD of the measured laser intensity in one experiment run of 11.6 days is as narrow as the detection bandwidth of a Discrete Fourier Transform $\delta f = \frac{1}{t} = 1\,\mu\text{Hz}$. Hence, we calculate the corresponding $i(\omega)$ at angular frequency $\omega$ by

$$i(\omega) = \frac{1}{t}\left|\int i(t)e^{i\omega t}dt\right|. \tag{16}$$

We fine-tune the frequency $\omega/2\pi$ every $0.1\,\mu\text{Hz}$ to find the maximum of $i(\omega)$, and set this frequency as the measured drive frequency $\omega_{\text{dri}}/2\pi$, as shown in Extended Data Fig. 5b. For one experiment run, indeed, we get $i(\omega_{\text{dri}}) = 0.317 = 0.995 \times 1/\pi \simeq 1/\pi$. We find that, the calculated $i(\omega)$ of the measured laser density is consistent with an ideal square wave signal that has the same amplitude and the same period of $2\pi/\omega_{\text{dri}}$. Hence, we treat $i(t)$ as an ideally periodic signal at frequency $\omega_{\text{dri}}/2\pi$ in our experiment.

**Noise calibration**

The displacement of motion is denoted as $X(t)$. We record the voltage output from the photodiode. To transfer the measured voltage into displacement, we first set the pressure of the vacuum chamber to high pressure ($4 \times 10^{-2}$ mbar) and calculate the effective temperature $T_{\text{eff}}$ from the measured PSD of oscillator's displacement using $k_B T_{\text{eff}} = m\omega_0^2/2\pi \int S_X(\omega) d\omega$. Under such a condition, the oscillator is fully thermalized and therefore the thermal Brownian motion satisfies the equipartition theorem $m\omega_0^2/2\pi \int S_X(\omega) d\omega = k_B T_{\text{en}}$, that is

$$T_{\text{eff}} = T_{\text{en}} \tag{17}$$

Extended Data Fig. 6a plots the PSD of the corresponding voltage of photodiode $S_V(\omega)$, where the following relation has been used[28]:

$$\int S_V^{\text{tot}}(\omega) d\omega = \xi^2 \int S_X(\omega) d\omega + \int S_V^{\text{mea}}(\omega) d\omega \tag{18}$$

The scaling factor $\xi$ between the displacement $X(t)$ and the measured voltage $V(t)$ is defined as $V(t) = \xi X(t)$; and the term $S_V^{\text{mea}}(\omega)$ is PSD of measurement noise in unit of V$^2$/Hz. In the current system, the measurement noise is treated as white noise $S_V^{\text{mea}}(\omega) = S_V^{\text{mea}}$ in the frequency range of interest. The uncertainty of effective temperature $\Delta T_{\text{eff}}$ scales as:

$$\frac{\Delta T_{\text{eff}}}{T_{\text{eff}}} = \sqrt{\frac{2}{t\gamma}} \tag{19}$$

with $t$ being the measurement time, as shown in Extended Data Fig. 6b. Hence, we get the uncertainty of the scaling factor $\xi$, which is given in Extended Data Table 1.

In the "fifth force" measurement, the vacuum chamber is pumped to high vacuum ($10^{-6}$ mbar) to reduce the mechanical dissipation. Force noise level needs to be determined before the data process. The force noise level is characterized by the power spectral density:

$$S_F^{\text{tot}}(\omega) = 4m\gamma k_B T_{\text{eff}} + S_F^{\text{mea}}(\omega) \tag{20}$$

where $\gamma$ is the mechanical dissipation measured from the PSD fitting as described above. In an ideal condition where the system is in thermodynamic equilibrium, we should always have the relation $T_{\text{en}} = T_{\text{eff}}$. For the case of high vacuum, in practice, external noises such as the vibration from environment and laser heating, can lead to higher $T_{\text{eff}}$. While in the current system, such potential noises have been well suppressed. In our experiment, the measurement process lasts more than 23 days. Although we have employed a temperature control scheme, there are still some drifts in the scaling factor $\xi$ due to temperature drift. The optical measurement scheme is sensitive to temperature changes so that $\xi$ is temperature dependent. By using the temperature control scheme, such drifts are suppressed significantly. During the experiment, we calculated $\xi$ every 10000s and followed the above process to obtain $X(t)$. The relative uncertainty of effective temperature is less than 6% in the current experiment. And the relation $T_{en} = T_{eff}$ holds within the measurement error (see Extended Data Table 1).

**Experimental data processing**

The system is modeled using the standard equation-of-motion of a harmonic oscillator:

$$m\ddot{X}(t) + m\gamma\dot{X}(t) + m\omega_0^2 X(t) = F_{th}(t) + F_{dri}(t) \tag{21}$$

with $X(t)$ being the displacement, $m$ the mass of the oscillator, and $\omega_0$ the resonance angular frequency. The dot denotes the time derivative with respect to $X(t)$. The first term on the right-hand-side is the stochastic thermal Brownian noise which satisfies the relation:

$$\langle F_{th}(t)F_{th}(0)\rangle = 4m\gamma k_B T_{eff}\delta(t). \tag{22}$$

The second term is the periodic drive force generated by the source masses:

$$F_{dri}(t) = F_{cham}(t) + F_{mag}(t) + F_{grav}(t) + F_{ele}(t) + \cdots \tag{23}$$

where $F_{cham}(t)$ is the fifth force from chameleon field that is of interest, $F_{mag}(t)$ is the magnetic force, $F_{grav}(t)$ is the Newtonian gravity, $F_{ele}(t)$ is the electrostatic force, and other inexplicit terms are negligible and omitted. We set the drive frequency $\omega_{dri}/2\pi$ at the resonance frequency $\omega_0/2\pi$ of the force sensor along the z-direction. There would also be high harmonics, which generate ignorable excitations at multiples of frequency $\omega_0/2\pi$.

In our measurement, both the rotation of source masses $i(t)$ and the motion of the force sensor $X(t)$ are record simultaneously. We calculate the corresponding displacement response

$X(\omega_{\text{dri}})$ at the drive frequency by

$$X(\omega_{\text{dri}}) = \frac{1}{t}\left|\int X(t)e^{i\omega_{\text{dri}}t}dt\right| \tag{24}$$

with $t$ being the measurement time. We denote $X_{\text{near}}(\omega_{\text{dri}})$ as the total displacement measured at the near position, which contains effects of both thermal fluctuations and drive forces from the source masses:

$$\langle X_{\text{near}}^2(\omega_{\text{dri}})\rangle = \langle X_{\text{th}}^2(\omega_{\text{dri}})\rangle + \langle X_{\text{dri}}^2(\omega_{\text{dri}})\rangle \tag{25}$$

Here, $X_{\text{th}}(\omega_{\text{dri}})$ corresponds to the response of mechanical oscillator to thermal noise with source masses being moved far away: $\langle X_{\text{far}}(\omega_{\text{dri}})\rangle = \langle X_{\text{th}}(\omega_{\text{dri}})\rangle$. When the drive frequency $\omega_{\text{dri}}/2\pi = \omega_0/2\pi$, $X_{\text{th}}(\omega_0)$ has an average $\langle X_{\text{th}}(\omega_0)\rangle = \sqrt{\frac{4k_B T_{\text{en}}}{m\omega_0^2 \gamma t}}$ for $t \gg 1/\gamma$. We then estimate the upper bound of $X_{\text{dri}}$ by using the relation:

$$[X_{\text{dri}}^{\text{up}}(\omega_{\text{dri}})]^2 = \max\left\{\left(X_{\text{near}}^2(\omega_{\text{dri}}) - X_{\text{far}}^2(\omega_{\text{dri}})\right), 0\right\} + 2\sqrt{2}\sigma^2[X_{\text{far}}(\omega_{\text{dri}})] \tag{26}$$

which corresponds to a 95% confidence level. $X_{\text{near}}(\omega_{\text{dri}})$ and $X_{\text{far}}(\omega_{\text{dri}})$ are measured in two independent experimental runs at near and far positions, respectively. The corresponding standard deviation $\sigma[X_{\text{far}}(\omega_{\text{dri}})]$ is calculated from the measured data $X_{\text{far}}(\omega_{\text{dri}})$.

Then we calculate the upper bound of the drive force $F_{\text{dri}}(t)$. To this end, we firstly define

$$F_{\text{dri}}(\omega) = \frac{1}{t}\left|\int F_{\text{dri}}(t)e^{i\omega t}dt\right| \tag{27}$$

The upper bound $F_{\text{dri}}^{\text{up}}(\omega_{\text{dri}})$ of the drive force $F_{\text{dri}}(\omega_{\text{dri}})$ is then calculated by:

$$F_{\text{dri}}^{\text{up}}(\omega_{\text{dri}}) = \frac{mX_{\text{dri}}^{\text{up}}(\omega_{\text{dri}})}{|\chi(\omega_{\text{dri}})|} \tag{28}$$

with $\chi(\omega) = 1/(\omega_0^2 - \omega^2 + i\omega\gamma)$ being the mechanical response of the mechanical oscillator. For the source mass driven by a servo motor, we get the periodic "fifth force" from numerical simulations of chameleon potential at different phases $F_{\text{cham}}(\theta)$ with $i(t)$ offering the drive angular velocity $\omega_{\text{dri}}(t)$:

$$F_{\text{cham}}(\theta(t)) = F_{\text{cham}}\left(\int \frac{1}{8}\omega_{\text{dri}}(t)dt\right) \tag{29}$$

During the whole experiment, the linewidth of drive is negligible as described above, so that $\omega_{\text{dri}}(t)$ can be treated as a constant, and we get $F_{\text{cham}}(\omega_{\text{dri}})$ from Eq. (14):

$$F_{\text{cham}}(\omega_{\text{dri}}) = \eta F_{\text{cham}}^{\text{pp}} \tag{30}$$

Here $\eta$ is a geometrical factor calculated from the normalized phase function $f(\theta)$, which is only weakly dependent on the Compton wavelengths and is regarded as a constant $\eta = 0.315$. The uncertainty induced this way is $\lesssim 0.5\%$. $F_{\text{cham}}^{\text{pp}}$ is the peak-to-peak value of $F_{\text{cham}}(\theta(t))$ during one period. Hence, we get the upper bound of $F_{\text{cham}}^{\text{pp}}$:

$$F_{\text{cham}}^{\text{up}} = \frac{1}{\eta} F_{\text{dri}}^{\text{up}}(\omega_{\text{dri}}) = \frac{m X_{\text{dri}}^{\text{up}}(\omega_{\text{dri}})}{\eta |\chi(\omega_{\text{dri}})|} \tag{31}$$

**Background forces and the shielding scheme.**

There are three main coherent background forces generated by the source masses with similar time dependence as the "fifth force". They are magnetic, electrostatic forces and Newtonian gravity. The magnetic force is mainly due to the magnetic field $\boldsymbol{B}_{\text{ba}}(t)$ that is induced by the source films with non-zero magnetic susceptibility $\chi_{\text{film}}$ in the presence of non-zero background magnetic fields mainly from the magnets. The magnetic force $\boldsymbol{F}_{\text{mag}}(t)$ can be expressed as

$$\boldsymbol{F}_{\text{mag}}(t) = \iiint_V -\nabla U(\boldsymbol{X}, t) \, dV \tag{32}$$

$V$ is the volume of the force sensor and $U(\boldsymbol{X}, t)$ is the total potential energy density of the magneto-gravitational trap[45]

$$U(\boldsymbol{X}, t) = \frac{|\chi|}{2\mu_0} \big(\boldsymbol{B}_0(\boldsymbol{X}) + \boldsymbol{B}_{\text{ba}}(t)\big)^2 \tag{33}$$

with $\boldsymbol{B}_0(\boldsymbol{X})$ being the time independent magnetic field of the magnetogravitational trap, and $\chi$ being the magnetic susceptibility. In the experiment, we measured the magnetic susceptibility of the film $\chi_{\text{film}}$ with a Superconducting Quantum Interference Device (SQUID), which is given in Extended Data Table 1. To suppress such magnetic forces, a magnetic shielding box is used to minimize the magnetic field generated by the magnetogravitational trap and the induced magnetic field $\boldsymbol{B}_{\text{ba}}(t)$ in the regions of the source masses. Extended Data Fig. 7 shows the distribution of magnetic fields and the estimated magnetic forces in one period.

A similar magnetic shielding box is also used to enclose the servo motor. As such, the magnetic field generated by the servo motor is suppressed. In practice, the shielding box is made of 1 mm-thick permalloy. A small hole of diameter 2 mm is left at the top of the shielding box so that the test mass of the force sensor can get out of the box.

The second background force is the electrostatic force. The charges on the source masses and the test mass can generate Coulomb forces. We eliminate such forces by using electrical shielding. In our experiment, the electrical shield consists of the vacuum chamber and a metalized thin film, which is at the top of the chamber between the source masses and the force senor. Since the shielding film can also shield the chameleon field, we make it as thin as possible so that such shielding effect is minimized for the chameleon field in the parameter space of interest. In practice, we choose a silicon nitride window with a thickness of 400 nm and stress less than 250 MPa. The silicon nitride window is strong enough that it can sustain the pressure between the atmosphere and vacuum. To achieve electrical shielding, an aluminum layer of about 100 nm is evaporated onto the silicon nitride film. The silicon nitride window chip is then fixed on the vacuum flange using vacuum epoxy glues. Silver paint is employed to connect the chip and the steel flange so that the silicon nitride window and the vacuum chamber become fully equipotential. The whole system then can be treated as an ideal Faraday cage. As such, the electrostatic field generated outside the chamber can be eliminated. Extended Data Fig. 8 shows the simulated distribution of electric fields.

The third background force is Newtonian gravity of the source masses themselves. To minimize such effects, we have employed a thin-film structure instead of a bulk structure for the source masses. The effective mass of the source masses is about 3.1 milligram. The corresponding Newtonian gravity force $F_{\text{garv}}(\theta)$ can be directly calculated using Newton's law of gravitation, as shown in Extended Data Fig. 9. In current system, the Newtonian gravity is over two orders of magnitude below the expected "fifth force" of the chameleon field in the parameter space of interest.

**Acknowledgements** We thank Di Wu for the measurement of material magnetic susceptibility and Xing Rong for helpful discussions. The silicon nitride window was fabricated in the microfabrication center of the National Laboratory of Solid State Microstructures (NLSSM) and the numerical calculations were performed on the computing facilities at the High Performance Computing Center (HPCC) of Nanjing University. This work was supported by the National Key R&D Program of China (Grant No. 2018YFA0306600), the National Natural Science Foundation of China (Grants No. 12075115, No. 12075116, and No. 11890702), the Chinese Academy of Sciences (Grants No. XDC07000000 and No. QYZDY-SSW-SLH004), and Anhui Initiative in Quantum Information Technologies (Grant No. AHY050000).

**Author contributions** All authors contributed to the development and writing of this paper.

    P.H., J.H., J.D. conceived the general idea.

    R.L., P.H., P.Y., J.D. designed and built the experimental setup.

    P.Y., X.B., H.X. performed the measurements.

    C.Y., X.X., J.H. performed the theoretical calculations.

    J.H., P.H., J.D., P.Y., C.Y., C.-K.D. wrote the manuscript.

    P.H., J.H., J.D. supervised the research.

**Competing interests** The authors declare that they have no competing financial interests

**Correspondence and requests for materials** should be addressed to hp@nju.edu.cn (P.H.); hejianhua@nju.edu.cn (J.H.); djf@ustc.edu.cn (J.D.)

**Data and Code Availability**. All data needed to evaluate the conclusions in the paper are presented in the paper or the supplementary materials. The code for numerical simulations underlying this article will be shared on reasonable request to the corresponding authors.

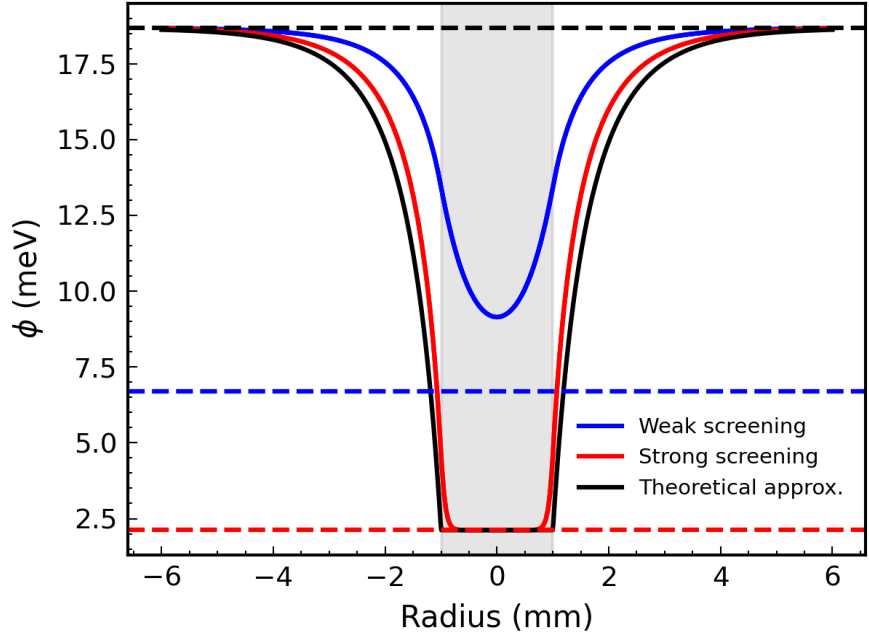

**Extended Data Fig. 1 | Numerical tests of the chameleon field for a sphere surrounded by air.** The horizontal axis shows the position relative to the spherical center, and the vertical axis is the magnitude of the field. The sphere is of 1 mm in radius (gray shaded regions). Blue and red curves show the results for the weak ($\rho = 10$ kg/m$^3$, $r_c = 0.27$ mm) and strong screening ($\rho = 100$ kg/m$^3$, $r_c = 0.05$ mm), respectively. Dashed horizontal lines indicate the equilibrium value of the field $\phi_{eq}$ for the air and sphere, respectively. The black curve shows the theoretical approximation[42] of the field under the strong screening case.

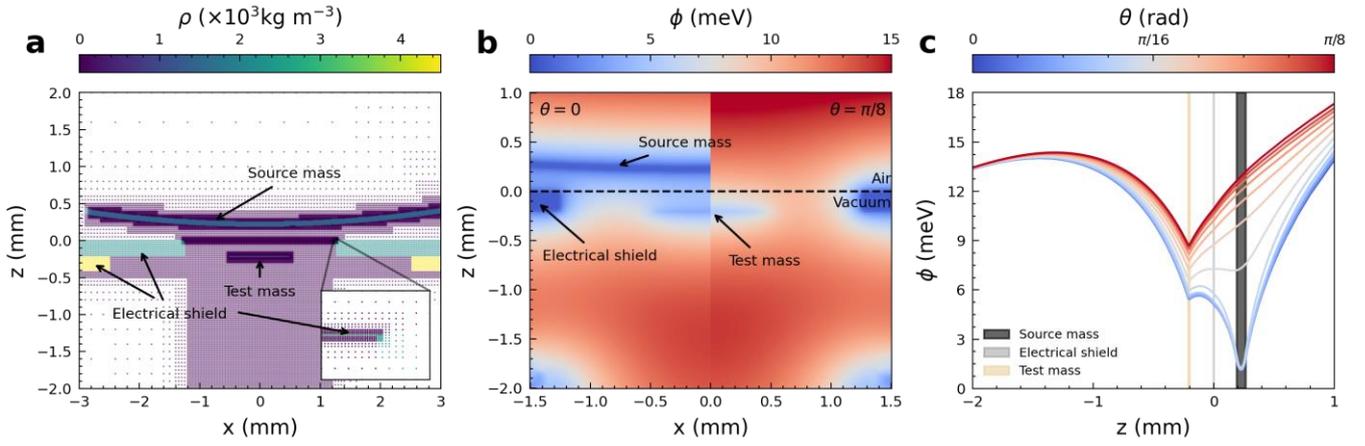

**Extended Data Fig. 2 | Numerical calculation of the chameleon field in our experiments.**
**a**, Non-uniformly refined grid points (DoFs) used for the FEM calculation. Different colors represent the matter density of the instrument with a color-bar at the top. The inset shows the grids around the electrical shield, where the finest resolution is about 0.5 $\mu$m. **b**, The distribution of the chameleon field $\phi$ around the test mass of the force sensor with (left half, phase angle $\theta = 0$) and without (right half, $\theta = \pi/8$) the source masses. Here $\theta = 0$ is illustrated by Fig. 1b in the main text. **c**, The distribution of $\phi$ along the central z-axis at different phases from $\theta = 0$ (blue curves) to $\theta = \pi/8$ (red curves).

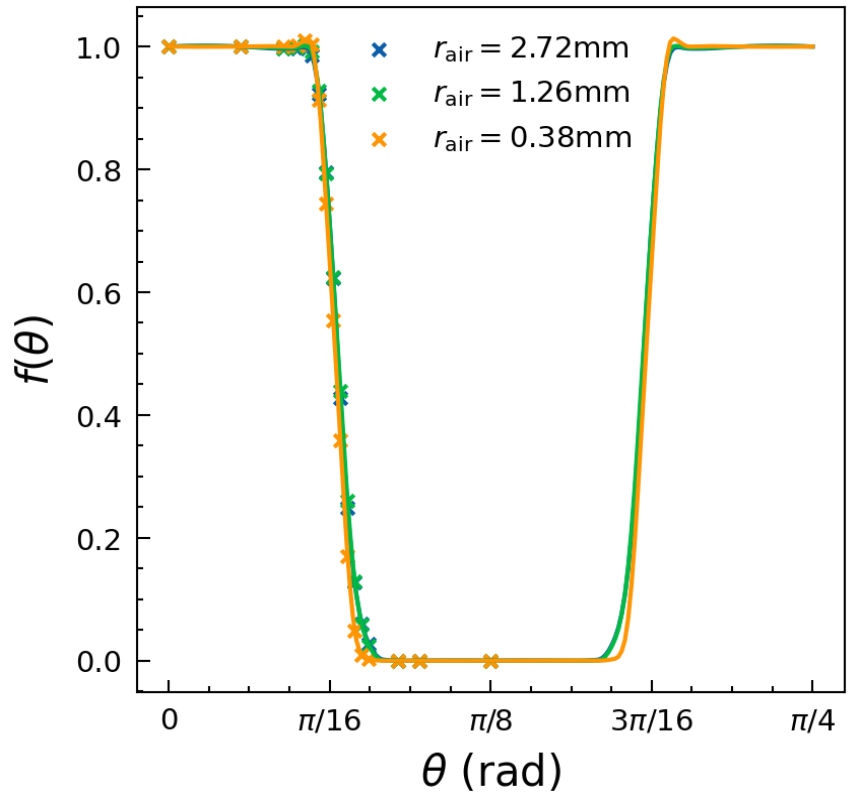

**Extended Data Fig. 3 | The normalized phase function of "fifth force" $f(\theta)$ felt by the force sensor in one period.** The marks are for the actual calculations under different phases (phase starts at the position as shown in Fig. 1b in the main text and rotates counter-clockwise). Solid curves are interpolations with quadratic splines. Different colors are for the models with different Compton wavelengths in the air.

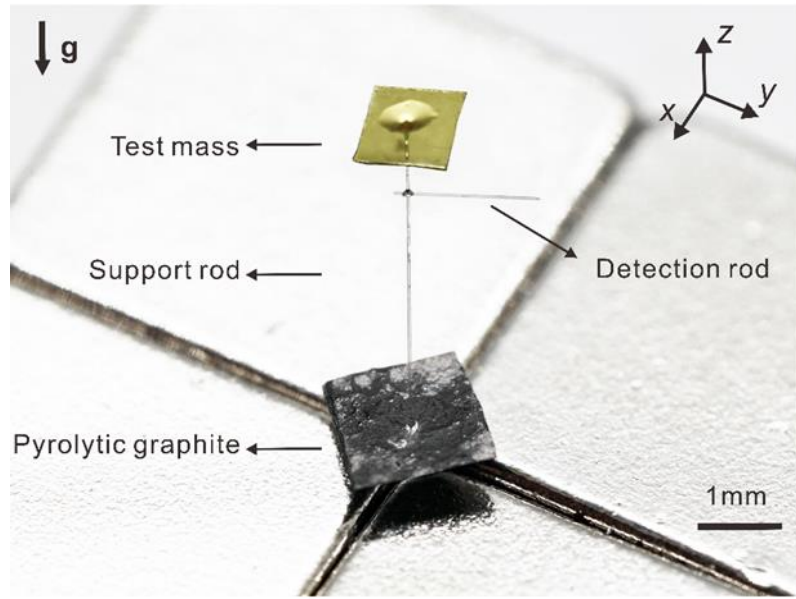

**Extended Data Fig. 4 | A representative diamagnetically levitated oscillator used as the force sensor in our experiment.** Gravitational force is along the negative z-direction. A thin film of polyimide with thickness of 12.5 um used as a test mass is fixed at the top of the force sensor, which is supported by the support rod; a 1.5 mm-long glass rod is used as the detection rod, which is attached horizontally on the support rod; a piece of pyrolytic graphite is used as the supporter, which is levitated above the magnets due to its strong diamagnetism.

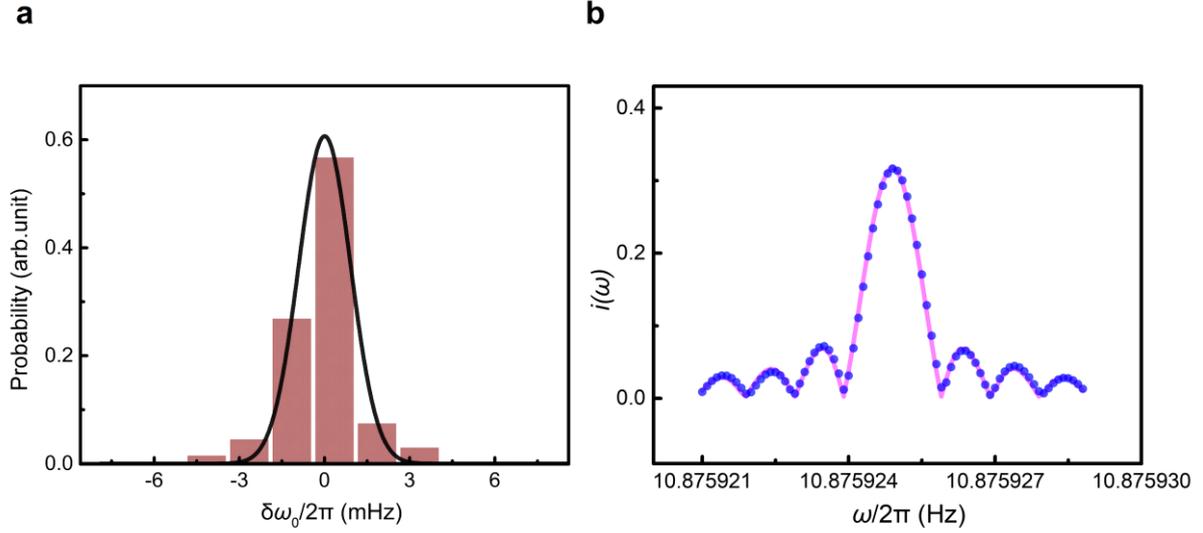

**Extended Data Fig. 5 | Frequency stability of the force sensor and the linewidth of source masses. a**, The measured distribution of the resonant frequency drift $\delta\omega_0/2\pi$. A single measurement of frequency is obtained from the fitted PSD of 4000 s. The measurement is fitted to a Gaussian distribution $P(\delta\omega_0)\sim e^{-\frac{(\delta\omega_0)^2}{2\sigma^2}}$. Here we obtain $\sigma/2\pi = 0.91$ mHz and $\omega_0/2\pi = 10.877$ Hz. **b**, The measured $i(\omega)$ as a function of the frequency $\omega/2\pi$. The blue dots indicate the measured laser intensity $i(\omega)$ for every 0.1 μHz. The magenta line shows the calculated $i(\omega)$ of an ideal square wave signal with the same amplitude and period. The maximum of $i(\omega)$ is $i(\omega)_{\max} = i(\omega_{\text{dri}}) = 0.317$, corresponding to the measured drive frequency $\omega_{\text{dri}}/2\pi = 10.8759249$ Hz.

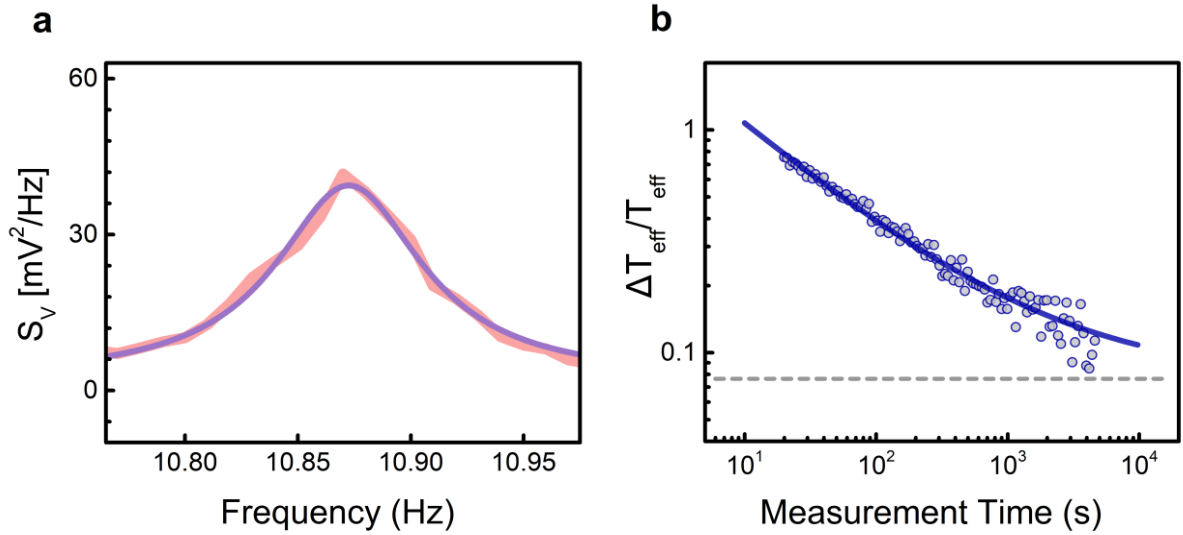

**Extended Data Fig. 6 | Thermal noise calibration. a**, The measured PSD of the voltage of photodiode at pressure $4 \times 10^{-2}$ mbar. **b**, The relative uncertainty of effective temperature as a function of measurement time $t$.

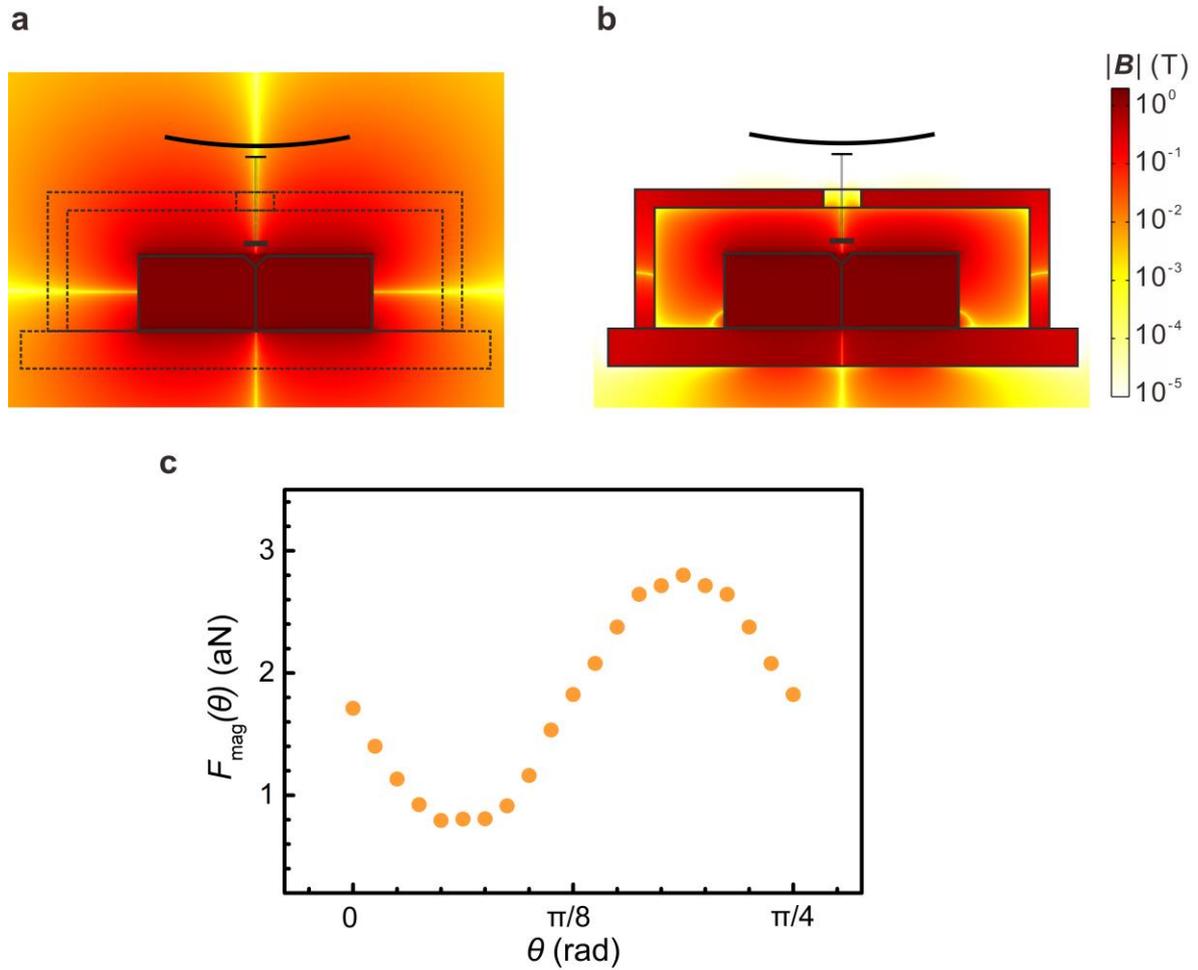

**Extended Data Fig. 7 | Numerical calculation of the magnetic force. a**, **b**, Magnetic field distribution without and with magnetic shielding box, respectively. In the numerical simulation, we calculated the magnetic field in regions including the main experiment set-up: source masses, magnetic shielding box, force sensor and the magnets. We set the rotation phase as $\theta = 0$, corresponding to the case with the source film on the top. **c**, Estimates of the magnetic force $F_{\mathrm{mag}}(\theta)$ induced by the rotating source masses as a function of $\theta$ in one period.

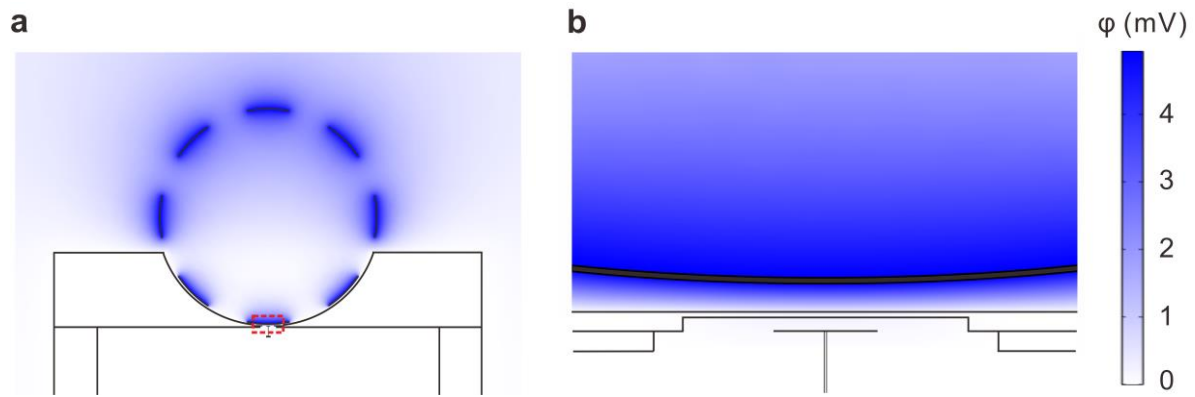

**Extended Data Fig. 8 | Numerical simulation of the electric field. a**, The electrical field distribution with an electrical shield consisting of a silicon nitride window chip and the vacuum chamber. In the numerical simulation, we calculated the electrical field in regions including the main experiment set-up of interest: source masses, electrical shield, and force sensor. We set $2 \times 10^4$ elementary charges on every source film and ground the electrical shield. **b**, Zoom-in of the area in the red dashed frame in **a**.

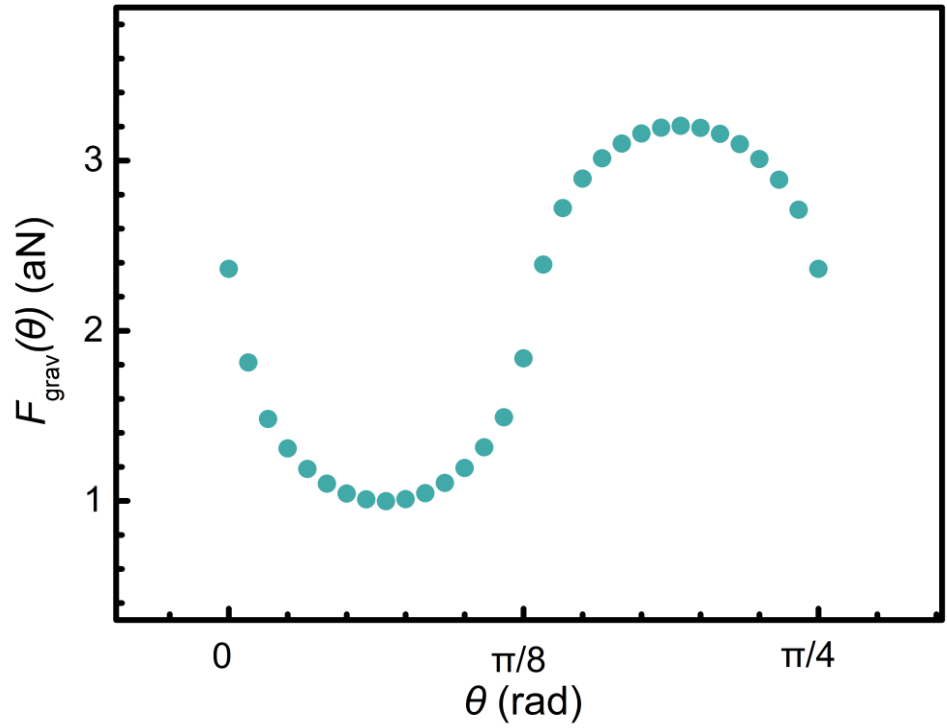

**Extended Data Fig. 9 | Numerical calculation of the Newtonian gravity.** Newtonian gravity $F_{\text{grav}}(\theta)$ generated by the rotating source masses in one period. We calculate $F_{\text{grav}}(\theta)$ according to Newton's law of gravitation.

# Extended Data Table 1 | Experiment parameters and deviation

| Symbol | Value | Unit |
| --- | --- | --- |
| $d_1$ | $200 \pm 10$ | μm |
| $d_2$ | $190 \pm 10$ | μm |
| $\alpha_x$ | $(0 \pm 5)°$ | |
| $\alpha_y$ | $(0 \pm 5)°$ | |
| $\alpha_z$ | $(0 \pm 10)°$ | |
| $\rho^*_{\text{film}}$ | 1.42 | g/cm$^3$ |
| $\chi_{\text{film}}$ | $(2.1 \pm 0.3) \times 10^{-5}$ | |
| $m$ | $348 \pm 2$ | μg |
| $\omega_0/2\pi$ | 10.877 | Hz |
| $\sigma/2\pi$ | 0.91 | mHz |
| $Q$ | 1099 | |
| $\xi$ | $1.3 \pm 0.07$ | $\times 10^6$ V/m |
| $\sqrt{S_X^{\text{th}}(\omega_0)}$ | 5.9 | $\times 10^{-9}$ m/$\sqrt{\text{Hz}}$ |
| $\sqrt{S_X^{\text{mea}}}$ | 1.3 | $\times 10^{-9}$ m/$\sqrt{\text{Hz}}$ |
| $\sqrt{S_F^{\text{th}}}$ | 19.0 | $\times 10^{-15}$ N/$\sqrt{\text{Hz}}$ |
| $\sqrt{S_F^{\text{mea}}(\omega_0)}$ | 3.7 | $\times 10^{-15}$ N/$\sqrt{\text{Hz}}$ |
| $\sqrt{S_F^{tot}(\omega_0)}$ | 18.8 | $\times 10^{-15} \text{N}/\sqrt{\text{Hz}}$ |
| $T_{en}$ | $298.24 \pm 0.008$ | K |
| $T_{eff}$ | $282 \pm 33$ | K |

* $\rho_{film}$ is the density of the thin films of source masses and force sensor, which are made of the same material polyimide.